\PassOptionsToPackage{dvipsnames,table,xcdraw}{xcolor}

\documentclass[acmsmall]{acmart}%,

\AtBeginDocument{%
	}

\usepackage{enumitem}
\usepackage{float}

\usepackage{multirow}
\usepackage{makecell}

\usepackage{csquotes}
\usepackage{amsmath}
\usepackage{bm}
\robustify\bfseries

\usepackage{siunitx}
\usepackage[position=top]{subfig}
\usepackage{graphicx}
\usepackage{pifont}
\newcolumntype{P}[1]{>{\centering\arraybackslash}m{#1}}
\newcolumntype{R}[1]{>{\raggedright\arraybackslash}m{#1}}
\newcolumntype{L}[1]{>{\raggedleft\arraybackslash}m{#1}}

\usepackage{xspace}

\usepackage{url}

\makeatletter
\renewcommand{\fps@figure}{htb}         % default {tbp}
\renewcommand{\fps@table}{htb}         % default {tbp}
\makeatother 

\usepackage[ruled]{algorithm2e}
\usepackage{cleveref}

\usepackage{longtable}

\newcommand{\FINAL}[1]{}

\newcommand{\transcript}[1]{\emph{``#1''}\xspace}
\newcommand{\interviewee}[1]{(\textsc{#1})\xspace}

\begin{document}

\title{A Qualitative Analysis of Remote Patient Monitoring: How a Paradox Mindset Can Support Balancing Emotional Tensions in the Design of Healthcare Technologies}

\author{Zoe Jonassen}
\affiliation{%
  \institution{School of Management, University of St. Gallen}
  \city{St. Gallen}  
  \country{Switzerland}}
\email{zoe.jonassen@unisg.ch}

\author{Katharine Lawrence}
\affiliation{%
  \institution{NYU Grossman School of Medicine}
  \city{New York}
  \country{United States}}
\email{katharine.lawrence@nyulangone.org}

\author{Batia Mishan Wiesenfeld}
\affiliation{%
  \institution{New York University}
  \city{New York}
  \country{United States}}
\email{bmw1@stern.nyu.edu}

\author{Stefan Feuerriegel}
\affiliation{%
  \institution{LMU Munich \& Munich Center for Machine Learning}
  \city{Munich}
  \country{Germany}}
\email{feuerriegel@lmu.de}

\author{Devin Mann}
\affiliation{%
  \institution{Department of Population Health, NYU Grossman School of Medicine}
  \city{New York}
  \country{United States}}
\email{devin.mann@nyulangone.org}

%\author{Author names blinded for review}

\renewcommand{\shortauthors}{Jonassen et al.}

\begin{abstract}
Remote patient monitoring (RPM) is the use of digital technologies to improve patient care at a distance. However, current RPM solutions are often biased toward tech-savvy patients. To foster health equity, researchers have studied how to address the socio-economic and cognitive needs of diverse patient groups, but their emotional needs have remained largely neglected. We perform the first qualitative study to explore the emotional needs of diverse patients around RPM. Specifically, we conduct a thematic analysis of 18 interviews and 4 focus groups at a large US healthcare organization. We identify emotional needs that lead to four emotional tensions within and across stakeholder groups when applying an equity focus to the design and implementation of RPM technologies. The four emotional tensions are making diverse patients feel: (i) heard vs. exploited; (ii) seen vs. deprioritized for efficiency; (iii) empowered vs. anxious; and (iv) cared for vs. detached from care. To manage these emotional tensions across stakeholders, we develop design recommendations informed by a paradox mindset (i.e., “both-and” rather than “and-or” strategies).
\end{abstract}

%%%
%%% The code below is generated by the tool at http://dl.acm.org/ccs.cfm.
%%% Please copy and paste the code instead of the example below.
%%%
%



%\received{July 2023}
%\received[revised]{October 2023}
%\received[accepted]{November 2023}

\maketitle

\section{Introduction} 

Remote patient monitoring (RPM) refers to the use of digital technologies to collect and analyze patient health data at a distance with the aim of improving the quality and accessibility of care \cite{Taylor2021,Thomas2021}. RPM draws upon data from digital technologies, including from wearable devices in widespread use (e.g., smartwatches and smartphones) to monitor patients’ vital signs and other physiological and behavioral data. By monitoring patients’ health status from a distance, healthcare professionals can then interact with patients through messaging, video calls, or in-person visits, and thus facilitate the early prevention and treatment of diseases such as diabetes or hypertension, as well as promote education and self-management among patients \cite{Thomas2021}. As an example, smartwatches have been found to be highly effective in detecting hypoglycemia as an acute complication of diabetes mellitus \cite{Maritsch2020}. Overall, in recent years, RPM use increased dramatically due to new reimbursement schemes \cite{Nixon2018} and the COVID-19 pandemic \cite{Mecklai2021}. 

RPM has the potential to foster digital health equity by improving care for diverse patient groups \cite{Mann2022,Taylor2021}.\footnote{Throughout our paper, we use the term ``diverse"' patient groups to refer to those with identities that go beyond the prototypical intersectional identity of an RPM end-user, (i.e., young, digitally-literate, White, male English speaker).}  RPM could make it considerably easier to access care for people of different ages, genders, ethnicities, and (dis)abilities \cite{Lawrence2023,Richardson2022}. For example, a large proportion of African Americans tend to live in under-resourced neighborhoods where access to healthcare organizations is often limited. Similarly, elderly patients often have mobility issues, which makes it challenging for them to travel long distances to visit physicians while at the same time needing frequent care for chronic conditions. Broadening RPM adoption is especially important given that the proportion of elderly people is rising in many parts of the world (e.g., 25\% of all US citizens will be over 60 by 2030) \cite{UnitedNations2015} and that there is an increase in age-related and often chronic diseases \cite{Lindsay2012}.

Despite potential benefits for health equity, RPM (like other digital health technologies) is not equally adopted throughout society \cite{Hsieh2008,Lyles2021,Sieck2021}. One of the principal reasons for this is that the needs of diverse patient groups are often neglected in the design of RPM health technologies \cite{Lawrence2023}. Such technologies are often developed for the ``digitally literate"' -- that is, ``patients may appear to be better candidates for remote monitoring if they are younger, technology-savvy, or are fluent English speakers"' \cite[p.\,7]{Sanders2020}. By means of focusing on young and technologically skilled users, digital tools like RPM may actually reinforce existing inequalities \cite{Crawford2020,Majchrzak2016,Richardson2022}.

Several studies have analyzed how users from diverse backgrounds interact with digital health technologies. The literature on digital health equity has repeatedly demonstrated that users from diverse backgrounds face larger challenges due to specific \emph{socio-economic factors} and \emph{cognitive needs} \cite{Burner2014,Chaudry2012,Crawford2020,Duque2019,Lyles2021,Richardson2022}. Yet, to the best of our knowledge, no prior study has analyzed the role of the \emph{emotional needs} of diverse patients during RPM design and implementation. Similarly, existing studies of empathy and human-centered design for vulnerable groups discuss the benefits and challenges when designers empathize with diverse users’ cognitive and emotional needs \cite{Bennett2019,Lindsay2012,Wright2008}. However, how emotional needs of diverse patients evolve and are addressed by designers and other stakeholders beyond the design phase and during subsequent implementation remains unclear. 

Neglecting emotional needs is problematic because experiencing positive emotions is a core human need and critical for positive relationships such as those between patients and healthcare providers \cite{Barsade2002,Fredrickson2004}. Hence, we expect that user interactions with RPM will not only be driven by \emph{socio-economic} and \emph{cognitive needs} but also by \emph{emotional needs} throughout the design and implementation process of RPM. Beyond that, the implementation of RPM is a highly collaborative effort that affects a variety of stakeholders’ interests over time \cite{Berg1999,Richardson2022}. Paying attention to these competing demands is critical to successfully integrating patients’ emotional needs across design and implementation phases. To address these neglected but important issues, we aim to study the emotional needs, and, based on them, we aim to suggest design recommendations for RPM technologies that help improve health equity. We thus ask the following two \textbf{research questions}~(RQs):
\begin{quote}
\begin{itemize}
\item \textbf{RQ1:} \emph{Which emotional needs must be fulfilled for diverse users during the design and implementation of RPM?} \\
\item \textbf{RQ2:} \emph{What kinds of tensions emerge within and across stakeholders when addressing patients' emotional needs during the design and implementation of RPM?}
\end{itemize}
\end{quote}

\noindent
In this paper, we perform the first qualitative study exploring the emotional needs of diverse patients when interacting with RPM. To this end, we conducted interviews ($N = 18$) with a variety of stakeholders involved in the design and implementation of RPM at a large healthcare organization in the US that we refer to as Healthorg. We also held four focus groups with diverse patients ($N = 21$) and observed the RPM design and implementation in action (25 meetings each lasting around 60 minutes). Additionally, we analyzed relevant documents with design specifications regarding the RPM initiative at Healthorg. Throughout our analysis, we used thematic analysis \cite{Braun2006}, which is systematic and rigorous in detecting patterns of meaning in texts.

Our main results are as follows: we identify four key \emph{emotional needs} of diverse patients: feeling heard (E1), feeling seen (E2), feeling empowered (E3), and feeling cared for (E4). We further find that these emotional needs result in four \emph{emotional tensions}, which are evoked by conflicting demands within and across stakeholders (i.e., designers, operational staff, and providers) involved in the design and implementation of RPM. Specifically, we identify the following four \emph{emotional tensions}: making diverse patients feel heard vs. exploited (T1), making diverse patients feel seen vs. deprioritized for efficiency (T2), making diverse patients feel empowered vs. anxious (T3), and making diverse patients feel cared for vs. detached from care (T4). To manage these emotional tensions, we develop specific design recommendations around adopting a \emph{paradox mindset}.

\textbf{Contributions.} The findings highlight the importance of emotional needs and arising tensions around RPM. This study thereby extends existing work around on human-centered design and digital health equity in three ways:
\begin{itemize}
\item We find that addressing diverse patients’ emotional needs throughout the design and implementation process of RPM is critical to foster health equity. To the best of our knowledge, we are the first to show that diverse patients using RPM not only have socio-economic and cognitive needs but also \emph{emotional needs}, and to identify four such emotional needs.
\item We further identify four \emph{emotional tensions} that arise across stakeholder groups when applying an equity focus to the design and implementation of RPM technologies.
\item We offer tailored design recommendations to balance the above emotional tensions by adopting a paradox mindset.
\end{itemize}

\textbf{Practical implications.} To facilitate the design and implementation process of RPM and similar collaborative health technologies, our analysis suggests benefits of using a paradox mindset wherein stakeholders embrace emotional tensions, search for synergies, and see them as opportunities. In brief, one should look for ``both-and"' rather than ``either-or"' solutions, thereby integrating the emotional needs of diverse patients with other stakeholders’ demands. Informed by a paradox mindset, we suggest the following practices:
\begin{itemize}
\item To identify design requirements for diverse patients and balance potential emotional tensions, all relevant stakeholders, or at least representatives of each group (designers, operational staff, and providers), should be included in the design phase to ``empathize"' with diverse patients' needs.
\item To improve the actual implementation, designers, operational staff, and providers should employ a ``both-and"' strategy, not only catering to cognitive needs of patients and fulfilling their own demands, but also explicitly catering to patients’ emotional needs. For example, to help down-regulate patients’ fears and anxieties during RPM use, patients should be contacted via phone by operational staff or providers to allow them to talk through those concerns.
\item To foster continuous use of RPM technologies and avoid the outcome of diverse patients feeling removed from their care and missing the human touch with their providers, providers should again take a ``both-and"' approach and complement RPM with video and in-person visits.
\end{itemize}

\section{Related Work}

\subsection{Promoting Health Equity with RPM}

The US Centers for Disease Control and Prevention declared that integrating health equity considerations into medical programs is critical to foster public health \cite{DicentTaillepierre2016}. Yet, deficiencies in the level of received care and differences in the health status of diverse patient groups (e.g., patients of different age, race, and socio-economic status) relative to other patient groups remain severe \cite{Abookire2020,Crawford2020,Richardson2022}. For example, compared to White patients, African American patients in the US suffer from poorer health status, and elderly patients face ageist biases in their perceived ability to use health-related technologies \cite{Williams2015}. 

Health equity can be potentially improved -- to a large extent -- through RPM. RPM employs digital technologies to collect patient data such as vital signs, disease symptoms, and behavioral patterns (e.g., sleep, diet) through remote and often asynchronous means. For example, in RPM, data are typically collected through wearable devices or manually using peripheral tools \cite{Taylor2021}. Subsequently, the collected data are shared with healthcare professionals for data-driven assessments, enabling them to make personalized recommendations to patients regarding disease prevention and treatment \cite{Sanders2020}. RPM further facilitates educating and empowering patients through better self-management of their medical condition \cite{Taylor2021}. In particular, RPM technologies have the potential to provide diverse patient groups with continuous and high-quality care in their homes, regardless of their location \cite{Mann2022,Sanders2020}. 

\subsection{Needs of Diverse Patients Groups Using RPM}

The successful design and implementation of RPM must overcome several practical challenges to address the needs of diverse patient groups. First, there are needs related to socio-economic factors (e.g., limited digital access and internet connectivity) that, if not addressed, reinforce inequalities \cite{Richardson2022,Stowell2018}. One design strategy to address this need is to develop devices that do not require WiFi and use cellular instead of Bluetooth connectivity \cite{Richardson2022}. 

Second, there are specific cognitive needs of diverse patients. Examples include low digital literacy (i.e., limited skills relevant for using digital technologies) and negative attitudes toward technology \cite{Richardson2022}. Prior work makes several important design recommendations to foster digital health equity, such as adapting the design interface of health apps to the cognitive needs of low-literacy populations \cite{Chaudry2012} or personalizing the messaging for diverse groups so that they cue specific behaviors (e.g., medication reminders) \cite{Burner2014}.

Third, as we hypothesize later, there can also be emotional needs. For example, patients may experience fear about being monitored \cite{Mann2022}. However, the emotional needs of diverse patient groups in RPM use have remained unclear. 

\begin{table}[tbph]
\centering
\caption{Comparison of Related Work and our Contributions}
\label{tbl:rw}
\footnotesize
\begin{tabular}{>{\raggedright}p{2cm} p{3.5cm} p{4cm} >{\columncolor[HTML]{FFD965}} p{3cm}}
\hline
\textbf{Literature stream} & \textbf{Socio-economic and cognitive barriers to using digital health technologies}  & \textbf{Design recommendations to overcome barriers} & \textbf{Our contribution}\\
\hline
\textbf{Digital health equity} & 
\begin{tabular}[t]{>{\raggedright}p{3.5cm}@{}}
$\bullet$ Limited access to technology \cite{Richardson2022,Stowell2018} \\
$\bullet$ Limited digital access \cite{Lyles2021,Sieck2021} \\
$\bullet$ Exclusion from trials to test technology \cite{Isaacs2016} \\
$\bullet$ Negative beliefs about digital health devices \cite{Crawford2020,Richardson2022} \\
$\bullet$ Lack of digital health literacy \cite{Burner2014,Crawford2020,Duque2019,Lyles2021,Richardson2022} \\
$\bullet$ Lack of self-efficacy -- perceived inability to use the technology \cite{Burner2014,Richardson2022} \\
$\bullet$ Language barriers \cite{Chaudry2012,Richardson2022} \\
\end{tabular} 
&
\begin{tabular}[t]{>{\raggedright}p{4cm}@{}}
$\bullet$ Development of devices that do not require WiFi and use cellular instead of Bluetooth connectivity \cite{Richardson2022} \\
$\bullet$ Invest in community-based organizations or local partnerships to make devices freely available in underserved communities \cite{Richardson2022} \\
$\bullet$ Provide technological support \cite{Richardson2022,Sieck2021}  \\
$\bullet$ Involve people from vulnerable groups in leadership positions \cite{Crawford2020}  \\
$\bullet$ Co-design technology with diverse patients \cite{Richardson2022} \\
$\bullet$ Include data analysis and interpretation tools in the patient-user interface \cite{Richardson2022} \\
$\bullet$ Allow patients to actively approve all data transmitted to clinicians \cite{Richardson2022} \\
$\bullet$ Person-centered care and messaging \cite{Burner2014,Crawford2020} \\
$\bullet$ Adapt design to specific cognitive needs of low literacy populations (e.g., begin every task from the same location) \cite{Chaudry2012,Richardson2022} \\
$\bullet$ Invest resources in language-concordant interventionist \cite{Isaacs2016,Richardson2022} \\
$\bullet$ Ensure multi-lingual device interfaces \cite{Richardson2022}
\end{tabular} 
&
\begin{tabular}[t]{>{\raggedright}p{3cm}@{}}
$\bullet$ We find that, beyond \emph{socio-economic} and \emph{cognitive needs}, continuously addressing \emph{emotional needs} is critical to fostering use of RPM across diverse patients \\
$\bullet$ We suggest adopting a \emph{paradox mindset} where tensions are embraced and synergies within and across stakeholder groups are explored 
\end{tabular}
\\
\hline
\textbf{Empathy- and human-centered design around health technologies for vulnerable groups}
& 
\begin{tabular}[t]{>{\raggedright}p{3.5cm}@{}}
$\bullet$ Empathy can also distance designers from the lived experiences of vulnerable groups where designers may privilege their own interpretations \cite{Bennett2019} \\
$\bullet$ Through empathizing, vulnerable people might also feel an emotional burden \cite{Lindsay2012} \\
$\bullet$ Vulnerable patients may suffer from fatigue because research around their needs is done on very few patients \cite{Wadley2013} \\
\end{tabular} &
\begin{tabular}[t]{>{\raggedright}p{4cm}@{}}
$\bullet$ Empathize through a variety of means, including narratives, dialogue, perspective-taking to understand emotional responses \cite{Wright2008} \\
$\bullet$ Draw upon existing support group of vulnerable groups so that they feel more comfortable (e.g., engage caregivers of people suffering from dementia) \cite{Lindsay2012}
\end{tabular} &
\begin{tabular}[t]{>{\raggedright}p{3cm}@{}}
$\bullet$ Beyond understanding emotional tensions in the empathizing phase of designer–patient relationships, we find that it’s key to incorporate other stakeholders’ needs and address emotional tensions within and across stakeholder groups along the design journey of RPM
\end{tabular} \\
\hline 
\end{tabular}
\end{table}

Neglecting emotional needs of patients is problematic because particularly vulnerable patients often have low self-esteem, low digital literacy, and low self-efficacy around RPM technologies \cite{Richardson2022}, which increases the likelihood that they will experience stress and frustration. This is a critical barrier, because negative emotions lead people to remove themselves from the situation or assume a narrow mindset blocking the exploration of novel solutions \cite{Fredrickson2004,Seo2004}. These reactions hamper continuous technology use, which is particularly critical for remote patient monitoring \cite{Fredrickson2004}. We offer a comprehensive analysis of the emotional needs of diverse patients during the design and use of RPM below. 

\subsection{Human-Centered Design Approaches to Understand Emotional Needs Around Digital Health Technologies}

Human-centered design approaches, including participatory \cite{Lindsay2012} and experience-centered design \cite{Balaam2019}, discuss the importance of understanding human needs from a holistic perspective \cite{Lindsay2012,Muller1993}. Human-centered design approaches focus on understanding the user’s social context as well as their thoughts and emotions \cite{Gooch2018,Lindsay2012}. To this end, human-centered approaches often use qualitative research methods to analyze narratives or dialogues from diverse users’ perspectives, such as the perspectives of elderly or disabled patients, so that designers can gather insights into the emotional experiences of vulnerable groups around health technologies \cite{Bennett2019,Lindsay2012,Wright2008}.

One way to elicit emotional needs is through empathizing \cite{Bennett2019,Lindsay2012,Wright2008}. Here, the previous studies show that creating an affective partnership -- where designers both allow themselves to be affected by the emotions of the vulnerable group while also attending to differences between their own and the vulnerable group’s experience -- is important to truly understand their experience and stands in contrast to simple imposition of the designers’ perspective \cite{Bennett2019}. 

Besides showing the benefits of empathizing with users, studies have also highlighted liabilities of empathizing with diverse users’ cognitive and emotional experience when interacting with technologies. For example, empathy can also distance designers from the lived experiences of people with disabilities because designers may favor their own interpretations \cite{Bennett2019}. Asking vulnerable groups to share their experience can trigger negative feelings because of the burden they experience from being sick or vulnerable \cite{Lindsay2012}. 

Existing studies make important contributions by discussing the benefits and challenges of empathizing with diverse users’ cognitive and emotional needs and seeking to understand their holistic context \cite{Bennett2019,Lindsay2012,Wright2008}. However, the emotional needs of diverse patients during RPM use are unclear. A particular blind spot concerns how emotional needs evolve during the design and implementation beyond the empathizing phase. Please see \Cref{tbl:rw} for an overview of the existing literature.

\textbf{Research gap:} Developing RPM for diverse patient groups is an ongoing challenge. Existing studies have analyzed the \emph{socio-economic} and \emph{cognitive needs} of diverse patient cohorts during RPM use. However, their \emph{emotional needs} have remained unclear. Given that RPM is a highly collaborative and complex effort that affects a variety of stakeholders’ interests, there may be conflicts between patients' emotional needs and the competing demands of other stakeholders. To fill this gap, we are the first to analyze emotional needs related to RPM, the emotional tensions that emerge within or across stakeholder groups around RPM, and to develop tailored design recommendations for addressing these tensions.

\section{Method} 

\subsection{Overview of Research Approach}

To explore the emotional needs of diverse patients and the tensions that may arise around these emotional needs within and across stakeholders, we conducted a single-case longitudinal study of the design and implementation of RPM in a large healthcare organization (Healthorg) located in the US. Given the dearth of studies about the emotional needs of diverse patients regarding the design and implementation of RPM and the tensions that may arise around those needs, we chose a qualitative approach that allowed us to gather rich empirical data \cite{Langley1999,VandeVen1992}.

RPM is an important technology to study given its potential to bring care closer to diverse patients. By monitoring patients asynchronously and remotely, RPM may impede close relations between patients and providers. However, feelings of relatedness, which are central to human well-being, emerge only if humans feel understood and cared for by others \cite{LaGuardia2000}, motivating the importance of examining patients’ emotional experiences around RPM. 

\subsection{Study Context}

Healthorg is a leader in implementing telehealth, app-based healthcare services, and, in particular, RPM. Healthorg runs a network of hospitals as well as outpatient clinics and group practices providing outpatient care for a large and diverse patient population. This includes a large federally qualified health system consisting of community-based healthcare providers that receive federal funds to provide primary care services in underserved areas. During the 2017--2022 period, RPM was deployed across the ambulatory care practice network (including primary care, specialty medicine services, and surgery practices). RPM programs were targeted to the management of both acute care (e.g., COVID-19) and chronic diseases (e.g., diabetes, hypertension).

Since December 2017, Healthorg had placed 8,307 RPM orders for 4,656 patients, with the highest number being orders to monitor patients’ blood pressure and glucose levels. Of all enrolled patients, 45\% had submitted at least one data point. The number of orders continuously increased from 2019 (855 orders), to 2020 (2,084 orders), and 2021 (3,305). A total of 65.1\% of all orders were for patients who identify as female. The largest share of orders by age was for patients from 35 to 44 years old (23.8\%), followed by patients over 64 (19.2\%). Over half of the orders were for patients identifying as White (54\%), followed by those identifying as Black (16.7\%).

\subsection{Healthorg Approach to Designing and Implementing RPM Technologies}

Healthorg aimed to develop holistic solutions taking into account the larger social context of care \cite{Berg1999,Richardson2022} and involved different stakeholder groups that can be broadly categorized into four groups: patients, designers, operational staff, and providers. As Healthorg did not focus on any particular group of diverse patients (e.g., older patients or women), we also did not focus on a subgroup.

The RPM design and implementation process at Healthorg followed three phases: During phase 1 (2017--2021), designers sought to empathize with diverse patients and learn about their needs, experiences, thoughts, and emotions vis-a-vis RPM. In 2021, the designers conducted four focus groups with patients of diverse demographics (e.g., age, gender, ethnicity). During the second ``prototyping"' phase (2021--2022), the operational staff developed technical prototypes and solutions for how to implement RPM within existing workflows. In spring 2022, Healthorg entered the third phase, during which providers began to test the developed prototypes with patients. Although the process is presented here in a linear manner, it was highly iterative in practice.

\subsection{Data Collection}

We collected data through (1)~retrospective interviews with designers, operational staff, and physicians, (2)~prospective focus groups with patients, and (3)~real-time observations between patients, operational staff and physicians, as explained below. We combined multiple sources to gather in-depth empirical data and minimize retrospective bias. To further triangulate our insights, we also reviewed over 50 documents associated with the RPM program, including internal meeting notes, patient education material, and operational workflow diagrams.

Throughout the data collection, we prioritized diversity over representativeness by ensuring that we included informants with different characteristics and by employing a purposeful sampling strategy, which refers to including a few information-rich cases \cite{Patton2002}. Semi-structured focus groups and interviews were prioritized over completely structured ones to be able to adapt questions and follow topics that may arise during the conversation \cite{Gaskell2000}.

The author team is composed of five researchers -- two experts in medicine and health technologies, one in computer science, and two in psychology and technology management. The experts in medicine and health technologies are practicing clinicians who were part of the strategy team around RPM implementation, while the other three authors observed but were not part of the strategic initiative. This combination blends deeply embedded perspectives with more detached viewpoints. The authors who had an outside view analyzed the data and discussed their interpretations with the medical experts.

\subsubsection{Retrospective interviews}

Between February and June 2022, we conducted 13 formal interviews ($N = 13$) and 6 informal interviews ($N = 5$) with various key stakeholders involved in RPM design and implementation, including designers gathering and incorporating user needs ($N = 4$), operational staff developing the RPM prototype and planning the rollout ($N = 5$), and physicians testing RPM at their practices ($N = 9$). Of the 18 interviewees, 8 identified as male and 10 as female; 4 aged 20--29, 7 aged 30--39, 3 aged 40--49, and 4 aged 50--59. 

During these interviews, we asked participants about their roles, experiences, and decisions related to the RPM design and implementation process, as well as their perception of how RPM could potentially foster or hinder health equity. We further asked whether there would be anything they would do differently, and what health equity meant for them. All formal interviews lasted from 30 to 60 minutes and were transcribed.

\subsubsection{Prospective focus groups}

To gather patients’ different perspectives, the designers in the RPM implementation team invited diverse patients with respect to ethnicity (6 identifying as Black, 12 as White, 2 as Hispanic, 1 as Asian), gender (11 identifying as female, 10 as male), and age (2 patients between 20--29; 11 patients between 30--39; 8 patients between 40--59). We hosted four 90-minute focus groups via Zoom in June 2021. In total, $N = 21$ patients participated. 

As part of the focus groups, patients were asked to put themselves in the shoes of certain groups who face unique challenges around using RPM, including (1) adopters of early versions of RPM, (2) parents, (3) caregivers for other adults, and (4) people managing a chronic condition. The focus groups were led by two designers who were trained in user-centered design. For each session, we analyzed the recordings, transcripts, and all other relevant materials. 

\subsubsection{Real-time observations}

We observed the following implementation activities: (1) 25 weekly RPM team meetings (each lasting 1 hour), during which participating providers and operational staff reflected upon opportunities and challenges for the design and implementation of RPM and discussed next steps; (2) an RPM ``demo day"' where the technology was presented to providers, and general barriers and facilitators of its use in clinical practice were discussed; and (3) three patient--physician/patient--operational staff interactions when onboarding the first patients to RPM. The researcher on site took extensive field notes at all sessions.

\begin{figure}
\centering
\includegraphics[width=0.9\linewidth]{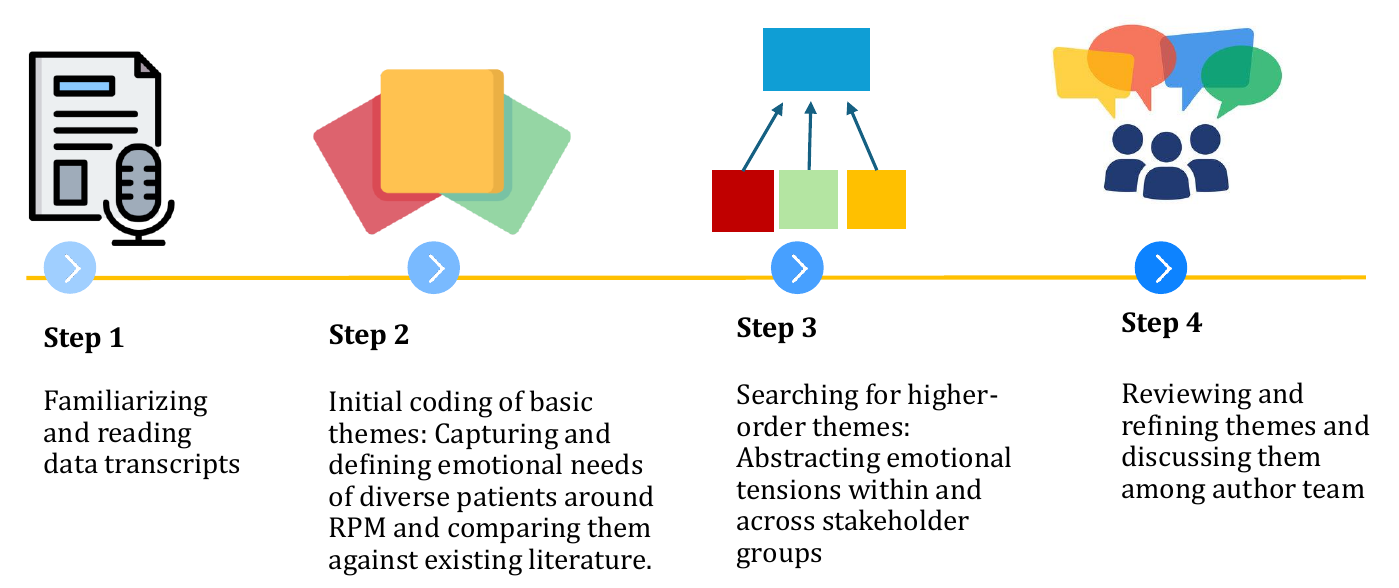}
\caption{Illustration of step-wise analysis process of the data}
 \Description{This image illustrates the four steps involved in the thematic analysis process for Remote Patient Monitoring (RPM): Step 1: Familiarizing and reading data transcripts. Step 2: Initial coding of basic themes, capturing and defining emotional needs of diverse patients around RPM and comparing them against existing literature. Step 3: Searching for higher-order themes, abstracting emotional tensions within and across stakeholder groups. Step 4: Reviewing and refining themes and discussing them among the author team.
 }
\label{fig:process}
\end{figure}

\subsection{Data Analysis}

The focus groups and formal interviews were our primary source of data. The number of participants is in line with prior studies using qualitative methods for CSCW \cite{Benk2022}. All our graphs are based on the 13 formal interviews and four focus groups. To code the transcripts, we selected a rigorous and explorative approach called thematic analysis (TA) \cite{Braun2006,Fereday2006}. TA allowed us to explore unexpected results as well as similarities and differences across stakeholders. Using the qualitative analysis software ATLAS.ti, we analyzed the data stepwise \cite{ATLAS.ti2022}. Please see \Cref{fig:process} as an illustration.

First, we familiarized ourselves with the data and re-read through the transcripts as soon as they became available \cite{Attride-Stirling2001}. In a second step, we began labeling the transcripts with short descriptions or words that summarized the met emotional needs and unmet emotional needs of diverse patients around RPM and that were close to the patient’s language (Please see an overview of the basic themes and their definitions in \Cref{tbl:coding_scheme}; Columns B and C). Hereby, we followed a hybrid approach of TA: all 11 basic themes were coded bottom-up (e.g., diverse patients fear losing close ``human touch"' in their relationships with their doctor). Some basic themes such as ``making diverse patients talk about their experience with the healthcare system and its technologies can be emotionally draining and burdensome due to past negative or traumatic experiences"' was in line with prior work suggesting that vulnerable people might also feel an emotional burden \cite{Lindsay2012}. We further highlighted links between emotional needs that were linked to prior research around cognitive aspects around RPM directly in \Cref{tbl:coding_scheme}; Column~C through adding the respective sources in brackets.

We then used structural codes to capture which voice the respective informant represented (patients, designers, operational staff, providers). During this process, we realized the prominent role of tensions in our data. These tensions were linked to conflicting needs of diverse patients in relation to RPM and emerged across groups of stakeholders who played a key role in the design and implementation of RPM such as designers, operational staff and providers. Re-reading the transcripts we realized that designers had internalized the voice of patients while operational staff had internalized the voice of providers. We reviewed the abstracted basic themes and distilled higher-order themes through revealing the tensions: (1) making diverse patients feel heard vs. exploited, (2) making diverse patients feel seen vs. deprioritized for efficiency, (3) making diverse patients feel empowered vs. anxious, and (4) making diverse patients feel cared for vs. detached from care. Please also see an overview of the revealed emotional tensions across stakeholder levels in \Cref{tbl:coding_scheme}; Column~A. The analytical process included going back and forth among different basic and higher order themes and merging them or further differentiating them. Finally, the identified themes were defined and refined through memo writing -- a function provided by ATLAS.ti \cite{ATLAS.ti2022}, which enables researchers to connect analytical thoughts with coded text segments. The emerging insights were repeatedly adapted by comparing them to existing literature and discussing them within the author team.

 \clearpage

\begin{center}
{\footnotesize
\begin{longtable}{>{\raggedright}p{2cm} >{\raggedright}p{2cm} >{\raggedright}p{2cm} p{6.5cm}}
%\centering
\caption{Overview of coding scheme for the thematic analysis.}
\label{tbl:coding_scheme} 
\\
%\tiny
%\begin{tabular}{>{\raggedright}p{2cm} >{\raggedright}p{2cm} >{\raggedright}p{2cm} p{6.5cm}}
\hline
Column A & Column B & Column C & Column D \\ 
\cmidrule(lr){1-1} \cmidrule(lr){2-2} \cmidrule(lr){3-3} \cmidrule(lr){4-4}
\textbf{Higher-order themes (RQ2)} & \textbf{Basic themes (RQ1)} & \textbf{Definition basic themes} & \textbf{Examples}\\
\hline
\hline
Diverse patients feel heard vs. exploited during design phase \textbf{(T1)}
& 
\textcolor{ForestGreen}{Need met ($+$):} Feeling heard through sharing rich stories during design phase of RPM design \textbf{(E1)}
&
Including diverse voices in development of technologies allows them to feel heard by sharing their own authentic experiences around RPM and their relationships with providers 
& 
\begin{tabular}[t]{>{\raggedright}p{6.5cm}@{}}
$\bullet$ \transcript{They [patients who participated in focus groups] were \ldots excellent, very diverse groups. We \textbf{paid a lot of attention to gender and race and age diversity} as we got our cohorts together, and the groups were phenomenal \ldots We literally had every kind of personality.} \interviewee{Interview, designer} \\ 
$\bullet$ \transcript{We had \textbf{some great insights} there [during the focus group with patients]. There was a Black woman who talked about how much better she liked her healthcare after she curated her providers to all be people of color \ldots She thought it was really important and was very difficult to find providers of color.} \interviewee{Interview, designer}
\end{tabular}
\\
\cmidrule(lr){2-4}
&
\textcolor{BrickRed}{Unmet need ($-$):} Sharing one’s story during design phase of RPM can be burdensome
&
Making diverse patients talk about their experience with the healthcare system and its technologies can be emotionally draining and burdensome due to past negative or traumatic experiences \cite{Lindsay2012}
&
\begin{tabular}[t]{>{\raggedright}p{6.5cm}@{}}
$\bullet$ \transcript{So, \textbf{having to deal with entrenched attitudes} and wider aspects of the healthcare delivery systems that \textbf{make it difficult for some people} certain populations are intermittently \textbf{engaged with the healthcare delivery} system because of the \textbf{emotional and psychological tax [they pay]} \ldots to engage with it [the system].} \interviewee{Interview, designer}\\
$\bullet$ \transcript{My own experience with my mom that interacted with the healthcare delivery system, experiencing racism vicariously through her as she interacted with the healthcare delivery system. She recently had a thyroid surgery, thyroidectomy. And again, \textbf{what she's had to deal with was very disrespectful} \ldots she still has to deal with very \textbf{disrespectful white male doctor being very aggressive} with her} \interviewee{Interview, designer, patient}
\end{tabular}
\\
\hline
Diverse patients feel seen vs. deprioritized for efficiency \textbf{(T2)}	
&
\textcolor{ForestGreen}{Need met ($+$):} Feeling seen through personalization of RPM \textbf{(E2)}
& 
Tailor care to specific needs of diverse patients \cite{Sanders2020}
&
\begin{tabular}[t]{>{\raggedright}p{6.5cm}@{}}
$\bullet$ Once the patient successfully connects her device to the RPM interface she smiles confidently and says: \transcript{Oh great. I think it (RPM) will be a wonderful opportunity. Particularly, when I am home, I \textbf{can just connect virtually whenever it suits me}.} \interviewee{Patient during our observations of one of the patient-operational staff interactions to onboard patient to use RPM} \\
$\bullet$ \transcript{It's (using RPM) still something that the \textbf{patient can do on their own time}, as opposed to having to take time off of work, travel to the doctor's office, get a single reading when they're potentially stressed out about all the logistics of getting to the doctor's.} \interviewee{Interview, physician}
\end{tabular} 
\\
\cmidrule(lr){2-4}
& 
\textcolor{BrickRed}{Unmet need ($-$):} Feeling neglected through prioritizing efficiency 
& 
Deprioritizing personalization of patient care to minimize extra work for providers through lean workflows
& 
\begin{tabular}[t]{>{\raggedright}p{6.5cm}@{}}
$\bullet$ \transcript{From the provider point of view, barriers are going to really be centered on \textbf{streamlining the workflow}, so how to \textbf{efficiently} and effectively \textbf{incorporate RPM} into their day-to-day processes.} \interviewee{Interview, designer}
$\bullet$ \transcript{They [providers] like to-the-point simplicity \ldots \textbf{Tools that will give you a capacity to focus on medicine}.} \interviewee{Interview, operational staff}
\end{tabular}
\\
\cmidrule(lr){3-4}
&
&
Selecting tech-savvy patients for RPM pilots instead of diverse patients where subgroups might lack tech-savviness
& 
\begin{tabular}[t]{>{\raggedright}p{6.5cm}@{}}
$\bullet$ \transcript{When we work with the clinicians, we try to give them what is the \textbf{patient criteria that is going to be most successful in this}? And often it is \textbf{meeting the clinical criteria}. But then also [the importance of being] relatively \textbf{tech savvy}, like he’s comfortable using an iPhone or an iPad.} \interviewee{Interview, operational staff} \\
$\bullet$ \transcript{I think \textbf{they need to be tech-savvy} in the sense that they understand the very, very basics. It’s not that it’s overly complicated but [that they] understand the basics.} \interviewee{Interview, operational staff}
\end{tabular}
\\
\hline
Diverse patients feeling empowered vs. anxious \textbf{(T3)} 
&
\textcolor{ForestGreen}{Need met ($+$):} Feeling empowered in care through using RPM \textbf{(E3)}
&
Sense of autonomy through taking control over their own health \cite{Taylor2021} 
&
\begin{tabular}[t]{>{\raggedright}p{6.5cm}@{}}
$\bullet$ \transcript{It allows them to monitor their conditions at home and to engage with clinical providers more often \ldots it just \textbf{empowers them to take control of their health} and \textbf{to be more proactive}. So, I think RPM is beneficial in that sense. Just in terms of consideration of minority patients.} \interviewee{Interview, designer} \\
$\bullet$ \transcript{What I’ve seen is through the app and \textbf{that control \ldots you have through remote patient monitoring}. A patient, by logging in their meals and seeing their own efforts through having their weight change through their own efforts that they’ve tracked, it’s almost a \textbf{positive feedback loop}.} \interviewee{Interview, designer} \\
$\bullet$ \transcript{They feel way more \textbf{encouraged from something they have done in seeing the results that they want to just keep it going} and have it become a lifestyle, which I think is a hope is that they'll want to do this self-sufficiently. To want to not only care about their health more, but also, \textbf{they feel like they want to keep going because they see the benefits they've done}.} \interviewee{Interview, designer} \\
$\bullet$ \transcript{You can get on a call with your doctor or \textbf{have your vitals checked through an app remotely} \ldots feel like you have control. Rather than feel this anxious feeling of, "Oh, I have no idea what's going on in my body.} \interviewee{Interview, designer}\\
$\bullet$ \transcript{So, you're \textbf{addressing things earlier on} rather than waiting it out and being like, "Maybe it's nothing." Then you go in and it's like, oh, it's stage four cancer or something. Yeah. So, I think being able to always think of your health remotely, not only makes you more cognizant of it.} \interviewee{Interview, designer} 
\end{tabular}
\\
\cmidrule(lr){2-4}
& 
\textcolor{BrickRed}{Unmet need ($-$):} Anxiety around technology and sense of losing control
& 
Diverse patients fear that that they are not capable of interpreting RPM data
&
\begin{tabular}[t]{>{\raggedright}p{6.5cm}@{}}
$\bullet$ \transcript{If you have \textbf{all of this data}, I feel like one might get kind of \textbf{nervous} about \ldots where do I fall within all of the data, is it normal.} \interviewee{Focus group 1, patient}\\
$\bullet$ \transcript{And it tells you how you slept and like your chatbot like sends you a message as soon as you're awake saying like: the perfect breakfast would include like carbs and like whatever, however many grams of protein or whatever \ldots \textbf{I get very lost in the data very quickly}.} \interviewee{Focus group 1, patient}
\end{tabular}
\\
\cmidrule(lr){3-4}
& &
Diverse patients feel helpless and overwhelmed by using RPM technology correctly
&
\begin{tabular}[t]{>{\raggedright}p{6.5cm}@{}}
$\bullet$ \transcript{I think there’s so much \textbf{fear around tech}, that \textbf{older people don’t want to actually use it}.} \interviewee{Interview, provider support staff} \\
$\bullet$ \transcript{And then there are also more intrinsic factors like self-efficacy -- \textbf{people may not feel as confident about using the technology} or being able to just manage dealing with other life stressors that they’re faced with. So, there’s a whole host of barriers and challenges that they have to deal with.} \interviewee{Interviewer, designer} \\
$\bullet$ During one of the training sessions with an operational staff, the patient fails to connect her device with the RPM interface and says to the operational staff member in a \textbf{slightly breaking voice}: \transcript{Oh, \textbf{I am not an expert in WiFi. It seems not to work}.} The \textbf{patient is looking around nervously}. She is breathing hectically. Her eyes are wide open and she looks at the operational support staff for help. \interviewee{Patients during our observations of one of the patient-operational staff interactions to onboard patient to RPM} 
\end{tabular}
\\
\cmidrule(lr){3-4}
& &
Diverse patients feel that they are losing control over their private life by being constantly monitored via RPM technologies \cite{Mann2022}
&
\begin{tabular}[t]{>{\raggedright}p{6.5cm}@{}}
$\bullet$ \transcript{A lot of our smart devices \ldots, step trackers or whatever, they’re giving us good information. But then there’s also a \textbf{risk that comes along with too much information}.} \interviewee{Focus group 3, patient}\\
$\bullet$ \transcript{I would just say I totally agree with (name of other focus group participant) about the \textbf{big brother}. A little bit intrusive. Especially the data collection and especially if you know someone's in your house that maybe doesn't want that done to them or maybe things are very private.} \interviewee{Focus Group 2, patient}
\end{tabular}
\\
\hline
Diverse patients feel cared for vs. detached from their providers \textbf{(T4)} 
& 
\textcolor{ForestGreen}{Need met ($+$):} Feeling cared for through immediate and continuity in care through RPM technologies \textbf{(E4)}
&
Patients receive timely and continuous feedback on their health status \cite{Mann2022,Sanders2020}
&
\begin{tabular}[t]{>{\raggedright}p{6.5cm}@{}}
$\bullet$ \transcript{It allows them to \textbf{engage with clinical providers more often}. Then also, [to] build maybe [a] stronger relationship with their provider by \ldots their \textbf{provider keeping track of their health} and them keeping track of their health and being able to speak about it and have those discussions, again from home.} \interviewee{Interview, designer} \\
$\bullet$ \transcript{I think one of the least attractive parts about visiting a doctor is the waiting \ldots Like, it's just the idea of sitting in a waiting room is literally painful for me to even think about. So, it seems like \ldots  the \textbf{saving of time is worth its weight in gold}.} \interviewee{Focus Group 2, patient} \\
$\bullet$ \transcript{I imagine kind of in your bedroom as you're winding down for the end of the day. You know, checking kind of where you landed with your vitals and whatever data and stats are being collected um in an app that's easily accessible that you \textbf{can kind of look at and say like, oh yeah, I did that and it's giving you the feedback to say like this is good}.} \interviewee{Focus Group 4, patient}\\
$\bullet$ \transcript{From a \textbf{provider perspective, they can see your vitals without directly}, really to talk with you and for the next visit, they would have all these information that can assist with the healthcare procedures and workflows. So, I think from that perspective, it \textbf{cuts down certain barriers between the patients and the provider}, at least that's my thought.} \interviewee{Interview, operational staff}
\end{tabular}
\\
\cmidrule(lr){2-4}
& 
\textcolor{BrickRed}{Unmet need ($-$):} Fear of losing human touch in relationship with their providers due to RPM
&
Diverse patients fear losing close human relationship with their provider by using RPM technologies
&
\begin{tabular}[t]{>{\raggedright}p{6.5cm}@{}}
$\bullet$ \transcript{But another thing about it though is just that we are still humans. I \textbf{feel like that face-to-face contact} \ldots, we \textbf{lose that sense} of, I don’t know, \textbf{connection} \ldots I can’t even say the word right now, but you lose that or \textbf{become detached from the doctor}.} \interviewee{Focus group 2, patient}\\
$\bullet$ \transcript{It’s like they [the doctors] \textbf{become so distant}, it’s like computers are taking over already, and \ldots \textbf{I still need to see my doctor}. I still feel like I want to go in, \ldots I want that communication. I need that even if it's not all the time, but I would still like to come into your office and speak to you face-to-face. Like maybe I need you to touch my arm and see if it’s okay.} \interviewee{Focus group 2, patient}\\
$\bullet$ \transcript{I guess \textbf{not losing the personal connection} of going to see the doctor in person where it's not always screen time. And um I thought that I mean I was nine months pregnant during Covid and my doctor wanted to do a tele med and I said you gotta be kidding me.} \interviewee{Focus Group 2, patient} \\
$\bullet$ \transcript{It's sort of it portends almost \textbf{a sort of dystopian}. You know we're going to look after you but from a distance you know it's not gonna be any hands-on doctoring here. It's all gonna be, you know, we'll give you this stuff you can check in online or whatever, \textbf{but you [get] no face to face from a provider}.} \interviewee{Focus Group 4, patient}
\end{tabular} 
\\
\hline
\end{longtable}
}
\end{center}

\subsection{Ethical Considerations} 

The data collection was part of a quality improvement and patient safety evaluation. Prior to collecting data, all researchers completed a quality improvement self-certificate approved by the Institutional Review Board at Healthorg. To ensure confidentiality, all transcripts were immediately anonymized. All interviews were preceded by an introductory phase during which the researcher explained the scope of the study and gave participants the opportunity to ask questions. Participants were further informed about their rights, including the opportunity to stop participation at any time, the confidentiality of their data, and the absence of risk associated with participating in the study. We also asked for their consent to participate and record the session.

Before the focus groups began, we further agreed on ground rules with patients such as welcoming all opinions, and that everyone should show respect for others by allowing any participant to finish their sentences. During the focus groups open dialogue was fostered by emphasizing that all voices were considered equally important, that there were no right or wrong answers, and that the information shared would be treated confidentially. At the end of the interviews and focus groups, we also offered participants the opportunity to add points that were not covered by facilitators or the interviewer.

Throughout the data collection and analysis, the research team took additional steps to protect informants’ privacy and safety. First, we shared data and derived findings only at the aggregate level. Second, we acknowledged that studying emotional needs can be a sensitive topic. To minimize the risk that informants felt uncomfortable, we did not deliberately ask them for their emotional needs but rather their general experiences regarding the design and use of RPM.

We engaged in reflection during the data collection process (e.g., evaluating whether questions required adaptations). Additionally, the researcher collecting the data reflected upon her own thoughts and feelings to minimize potential biases throughout the process. We further ensured that part of the team maintained a detached perspective and could play the role of a ``devil’s advocate"' regarding the conclusions we drew.

\section{Findings}

\subsection{Emotional Needs of Patients (RQ1)}

In this section, we provide data to describe emotional needs that must be satisfied for diverse patients regarding the design and use of RPM. Across different stakeholder groups involved in the design, implementation and use of RPM, we found that RPM potentially improves diverse patient experience through making them feel heard, seen with their unique needs, empowered to take control over their own health, and allowing them to feel cared for by their providers. Across stakeholder groups, feeling cared for and feeling seen were the most consistently prevalent emotional needs that could be fulfilled through RPM.

\subsubsection{Feeling heard (E1)}

Through being engaged in the co-design of RPM technologies through the focus groups, patients had the chance to share their voices, including their thoughts, fears and needs around RPM, and thereby feel heard. A rich story that was shared by a patient during one of the focus groups was RPM’s potential to make patients feel more comfortable and gather more accurate data. For example, one focus group patient explained that his mother usually reacts with strong anxiety when she interacts with doctors which elevates her blood pressure. The elevated blood pressure made doctors prescribe her medication. However, at home her blood pressure is at a normal rate. At the end of the focus group when the designers ask the patients whether they want to add something to the conversations, patients say that they have nothing to add and express that they felt that they could share all their thoughts and stories. Designers valued such rich stories from patients, as these allowed them to understand their experiences. For example, one designer who conducted the focus groups with diverse patients explained how the focus groups were extremely valuable for him to make design choices: \transcript{[the stories] can be informed by an anecdote from a particular person in a particular scenario \ldots Let’s say they’re caring for an elderly parent who doesn’t speak the same language and trying to manage their healthcare. [One wants] to be able to say \ldots ‘I’m actually going to address this real human need} \interviewee{Interview, designer}. By engaging closely with the patients during the focus groups, designers internalized patients’ experiences and voices.

\subsubsection{Feeling seen (E2)}

The personalized approach of RPM made patients feel seen, because through RPM their care became more personalized to their unique needs and lifestyles. As one patient explained: \transcript{There is a lot more personalization (...) which again makes it so much more about you.} \interviewee{Focus group 4, patient}. Similarly, one operational staff member emphasized RPM’s potential to foster equity through making patients feel seen in their unique needs: \transcript{[RPM technologies] provide a greater level of equity. You feel seen} \interviewee{Interview, operational staff}. One operational support staff member acknowledged that a number of diversity factors need to be addressed by RPM to be of use to diverse patients: \transcript{Like, not only genders are different, bodies are different, but also the color of your skin can mean vastly different medical conditions than it could for someone else} \interviewee{Interview, operational support staff}. Similarly, one of the designers emphasized the importance of making patients feel seen and to tailor RPM prototypes to specific patient groups’ needs early in the design process: \transcript{Yeah, I think solutions should be tailor-fitted to the respective groups  For instance, for black Americans who have more intrinsic factors like self-efficacy, people may not feel as confident about using the technology} \interviewee{Interview, designer}. One of the physicians also pointed out during a meeting that: \transcript{depending on which of the multiple set-ups you use, the on-boarding experience of patients is different depending on whether you onboard elderly people versus a 25-year-old} \interviewee{Observational notes RPM meeting}. 

\subsubsection{Feeling empowered (E3)}

RPM addressed patients' emotional need to feel empowered. RPM technologies enabled diverse patients to take control over their own health by actively measuring and tracking from wherever and whenever they wish and to ultimately make better decisions for their health based on the data. As one patient shared with us during the focus groups: \transcript{For me it's willpower, right? Mind over matter. For example, my company brings in pizza nearly every day, and there was a time when I was eating the pizza and I gained weight. I felt terrible \ldots I think you got to hold yourself accountable \ldots I don't need a coach to motivate me but I'm definitely interested in the app or the smartphone because I would like to get as much detailed information as possible.} \interviewee{Focus Group 1, patient}. Another patient shared this sense of increased autonomy with respect to their healthcare decisions: \transcript{It's like my personal choice, so that if those notifications are happening, I feel like I'm having to make the choice} \interviewee{Focus Group 1, patient}. Along similar lines, a technical support staff summarizing the diverse patient’s value of RPM as \transcript{being able to actually say, if their blood pressure seems a bit high, they can be more preventative with their health and keep track of it sooner} \interviewee{Interviewer, technical support staff}. 

\subsubsection{Feeling cared for (E4)}

Diverse patients reported that their need to feel cared for by their provider was met through RPM by receiving timely and continuous feedback on their health status. One focus group participant shared with us their appreciation for RPM notifying them about abnormal values: \transcript{[It indicates] this is in range, or you know that you should be concerned about [the reading], and you should reach out to your primary care physician when they open in the morning} \interviewee{Focus group 4, participant}. Patients further shared with us during the focus group that this immediate and constant care is particularly helpful for diverse patient groups such as the elderly because it counteracts their feelings of loneliness: \transcript{I know of a lot of elderly people who feel -- especially with the advent of the pandemic -- have felt very isolated \ldots Feeling like you have that kind of regular support [through RPM] could make all the difference in the world} \interviewee{Focus group 4, patient}.

\subsection{Emotional Tensions Within and Across Stakeholders to Access and Address Patients' Emotional Needs (RQ2)}

\begin{figure}
\centering
\includegraphics[width=0.9\linewidth]{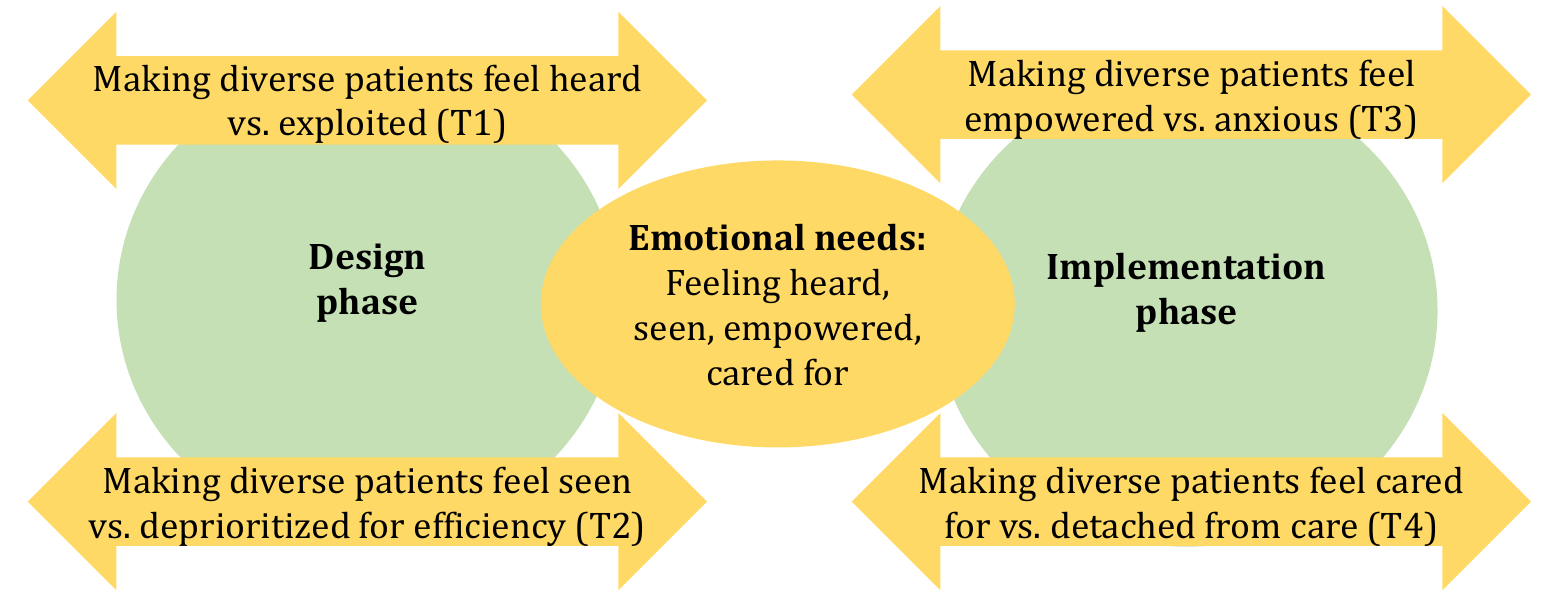}
\caption{Design and implementation process of RPM at Healthorg. Circle in the middle represents the emotional needs of diverse patients. Double arrows show emotional tensions that emerged in the design and implementation of RPM when addressing diverse patients’ emotional needs and other stakeholders’ demands.}
\Description{This image represents the emotional needs of diverse patients during the design and implementation phases of Remote Patient Monitoring (RPM): The central oval highlights the core emotional needs: feeling heard, seen, empowered, and cared for. The left green oval represents the Design phase, emphasizing: Making diverse patients feel heard vs. exploited (T1), and Making diverse patients feel seen vs. deprioritized for efficiency (T2). The right green oval represents the Implementation phase, emphasizing: Making diverse patients feel empowered vs. anxious (T3), and Making diverse patients feel cared for vs. detached from care (T4).  }
\label{fig:framework}
\end{figure}

\begin{figure}
    \centering
    \includegraphics[width=\linewidth]{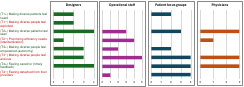}
    \caption{Mentioning of emotional needs and tensions per stakeholder, split into the four stakeholder groups. T1 to T4 represent the four tensions. The pluses and minuses behind the T's indicate whether a need was met versus unmet. For example, all four designers mentioned that diverse patients felt seen  (T2$+$), while only one of them mentioned that patient efficiency needs were prioritized (T2$-$).}
    \label{fig:stats_occurence}
    \Description{Stakeholder responses to emotional needs in RPM, categorized by Designers, Operational Staff, Patient Focus Groups, and Physicians.}
\end{figure}

\begin{figure}
    \centering
    \includegraphics[width=\linewidth]{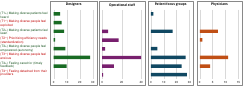}
    \caption{Frequency of emotional needs and tensions mentioned per stakeholder group. For example, patient focus groups mentioned T4$-$ a total of 28 times.}
    \label{fig:stats_freq}
    \Description{Stakeholder responses to emotional needs in RPM, categorized by Designers, Operational Staff, Patient Focus Groups, and Physicians.}
\end{figure}

In this section, we describe the emotional tensions that emerged around addressing the emotional needs of diverse patients during the RPM design and implementation process (RQ2). Please see \Cref{fig:framework} for an overview of our findings and \Cref{tbl:coding_scheme} for additional data examples. \Cref{fig:stats_occurence} and \Cref{fig:stats_freq} report the frequency, namely, the number of discussions mentioning the reported codes and the overall number of total mentions, thus underpinning the tensions per stakeholder group (patients, designers, operational staff, and physicians).

\subsubsection{Tension 1: Making diverse patients feel heard versus exploited (T1)}

When bringing patients’ emotional needs to light in the design phase, a key tension was experienced by designers between eliciting diverse patients’ voices about RPM and making them feel heard, versus the risk of exploiting patients when asking them to share personal experiences and views.

While the focus groups were highly valuable in accessing patients' experiences and rich stories, salient previous negative experiences with the healthcare system, including traumatic encounters, often overshadowed how diverse patients entered any relationship with the healthcare system. A designer who identified as African American pointed out that talking about experiences with the healthcare system can be emotionally challenging for many diverse patients because prior experiences are often negative. They observed: \transcript{So, I saw my mom experience racism with providers. And this happened to me when I was much younger -- it was very traumatic} \interviewee{Interview, designer}. One provider shared during an informal interview that providing input is fraught for such individuals and leaves them vulnerable to exploitation: \transcript{The issue is \ldots then if you are a Black woman, you are asked repeatedly to represent [African American women with] your voice. And you are not compensated for it! You have to do this on top of all other things that are going on in your life and this can be very tiring and emotionally draining} \interviewee{Informal interview, provider}.

To counteract diverse patients’ feelings of being exploited, designers applied a ``both-and"' strategy whereby they focused on both accessing diverse stories and minimizing possible feelings of exploitation. Designers were reflective and worked towards developing empathetic skills to not overburden patients during the focus groups. As one designer explained to us: \transcript{It's not an ad hoc process [to give voice to diverse patients]. We could fall back on having our accessibility expert to run everything. That's not fair. We should all learn how to do that. I should learn how to do it. Anybody working as a designer should be trained on doing an equity check of some sort} \interviewee{Interview, designer}.

In summary, designers perceive that accessing authentic life-experience stories by including diverse patients’ voices in the design process can become emotionally burdensome for patients who are asked to share deeply personal and often negative experiences. At the same time, accessing these real stories also makes diverse patients feel heard by allowing their early input in the process to inform later design choices. 


\subsubsection{Tension 2: Making diverse patients feel seen vs. deprioritized for efficiency (T2)}

A second emotional tension that emerged in the design phase was between addressing diverse patients’ emotional needs and healthcare staff demands for efficiency in the design of the prototypes. This tension was particularly experienced by operational staff members, because operational staff felt responsible for prioritizing simple and homogeneous solutions that ensure efficiency for providers. An operational employee pointed out that their team’s goal is to design the technology to fit existing workflows and thereby ultimately facilitate nurses’ and physicians’ use of RPM: \transcript{A lot of the decisions we make here are based on how it integrates with the EHR [electronic health record] system} \interviewee{Interview, operational manager}. Selecting solutions based on EHR integration constrained the options for addressing diverse patients’ emotional needs. As one physician concluded during one of the meetings: \transcript{What we have been thinking about is what tools do physicians need to deliver care? There is a lot of data that patients can send in. Who makes this decision what is sent in? Practices differ a lot. Sometimes nurses are overseeing it, sometimes not \ldots In the end it all comes down to cost issues} \interviewee{Observational notes RPM meeting}.

In hindsight, Healthorg decision-makers realized that they had neglected certain solutions that addressed diverse patients’ emotional needs to feel seen. 

One of the designers reflected that it might be more equitable in the future to move forward with multiple prototypes. This would allow them to explore which prototypes best address both diverse emotional needs and efficiency demands, and hence may lead to greater success with all groups. Through having been present during the focus groups with patients, designers embodied patients’ voices and felt a strong need to integrate them into the initial prototypes. As one designer explained: \transcript{Even if it’s a prototype, you’re already trying to make a good solution. It’s better to make three mediocre solutions and test those prototypes than [to] only solve parts of the problem} \interviewee{Interview, designer}. We observed that some stakeholders indeed began to develop a ``both-and"' approach, embracing and addressing both sides of the tension by experimenting with multiple pathways to determine which would best fit diverse patients’ emotional needs and users’ efficiency demands.

To summarize, the rich insights regarding distinctive users’ emotional needs gathered by the designers during the focus groups were at odds with the operational team’s directive to successfully meet efficiency requirements for physicians. This generated an emotional tension between stakeholder groups (i.e. patients and designers vs. operational staff and physicians) around building solutions that addressed diverse patients’ emotional needs of feeling seen versus building an integrated solution that optimized efficiency.

\subsubsection{Tension 3: Making diverse patients feel empowered vs. anxious (T3)}

New tensions emerged after the design phase when the RPM program entered the testing phase. Specifically, healthcare staff members had to balance the tension between making diverse patients feel empowered versus anxious when using RPM. While RPM was a means to empower patients to take control over their own health, engaging with the technology was also stressful and anxiety-provoking for diverse patients. Patients often did not feel capable of handling the autonomy that comes with RPM use and data interpretation and feared losing control over their data. For example, older patients felt that they lacked sufficient self-confidence to handle the data and their own healthcare decisions. Patients further shared with us their fears about being watched and controlled constantly. As one patient explained: \transcript{I don’t know that I need someone knowing everything about me at all times of the day just because I walked in my house. It may \ldots seem a little too big brother-ish, some aspects of these [technologies], particularly the data collection} \interviewee{Focus group 2, patient}.

An operational staff member who worked closely with older patients during the testing phase stressed that counteracting their feelings of helplessness and loss of control can be complex. Operational staff explained that over time they had learned how to address the emotional distress of diverse patients and prevent them from feeling overwhelmed and helpless. They now try to, for example, downregulate diverse patients’ fears due to low digital literacy by normalizing their experience and offering support: \transcript{A majority of the participants do not have a lot of tech literacy. So, I try to ease their minds by telling them ‘You’re not the only one’ and \ldots that ‘we are here to support you’} \interviewee{Interview, operational support staff}.

To summarize, our findings suggest that on the one hand, RPM has the potential to empower patients and provide them with autonomy to handle their health, while at the same time using and interpreting the RPM technologies can also be overwhelming and evoke feelings of anxiety and helplessness.

\subsubsection{Tension 4: Making Diverse Patients Feel Cared For vs. Detached from Care (T4)}

Another tension that came to the fore during the testing phase was between making diverse patients feel cared for versus making them feel detached from care when using RPM. While RPM allowed diverse patients to be more closely connected to their providers because of the continuous and immediate feedback through these technologies, diverse patients also reported feeling detached from their care due to missing the human touch. One patient, for example, shared with us: \transcript{I feel like with all these technological advances going on, I just don't want to lose that, kind of like everyone was saying, that human connection.} \interviewee{Focus Group 2, patient}. Similarly, another patient expressed: \transcript{Personally, the reason why I don't like these technology-mediated interactions is because sometimes you just want that interaction, you know what I mean? \ldots Sometimes I just want that smile to say everything's gonna be fine and stuff like that} \interviewee{Focus group 3, patient}. Some diverse patients feared losing their close relationship with their provider, as one patient explained: \transcript{Sometimes I think that when it comes to the personal care, the human touch seems to be what's been lost or glossed over.} \interviewee{Focus Group 4, patient}. Similarly, one designer explained: \transcript{Patients don’t want to be remote, and they don’t want to be monitored. They want to feel close to their care provider} \interviewee{Interview, designer}.

Instead of relying on technology-based interactions only, Healthorg began to complement RPM with calls by operational staff members and in-person visits with providers. In short, RPM has the potential to satisfy diverse patients’ need to feel cared for, which leads to a more equitable and proximate experience of care. However, diverse patients may then miss face-to-face contact with providers, all of which ultimately makes them feel detached from care. By combining remote care with both virtual and in-person visits, patients felt that they could maintain the ``human touch"' in their relationships with their doctor and experienced empathetic care based on someone in the healthcare system having firsthand knowledge of their personal history.

To summarize, while patients felt more closely connected to their healthcare provider through RPM, it also evoked feelings of detachment from personalized care and a loss of human connection with their providers.

\section{Discussion}

\subsection{Contributions}

Our work highlights the pivotal role of emotional well-being in technological design. Focusing on collaborative work, we reveal a set of emotional tensions that arise across stakeholder groups when applying an equity focus to the design and implementation of RPM technologies. Our study integrates and advances existing work on digital health equity \cite{Burner2014,Crawford2020,Lindsay2012,Massimi2007,Richardson2022} as well as on participatory \cite{Murray2023} and human-centered design \cite{Bennett2019,Lindsay2012}. Our work further links to and extends existing work around teams \cite{Harris2019}. More precisely, we extend existing these existing streams in three novel ways:
\begin{itemize}
\item We find that diverse patients have specific needs in regard to both healthcare technology and care in general. Previous work on RPM has revealed a set of socio-economic and cognitive needs of diverse patients’ using RPM \cite{Richardson2022}. To the best of our knowledge, we are the first study that \textbf{raises awareness of the importance of addressing diverse patients’ \emph{emotional needs}} around RPM. In particular, we identified four key emotional needs: \textit{feeling heard} (E1), \emph{feeling seen} (E2), \emph{feeling empowered} (E3), and \emph{feeling cared for} (E4). Being attentive to these emotional needs is critical because negative emotions can block critical processes needed for continuous RPM use, such as learning \cite{Shepherd2011}, and may spread across diverse patients and their communities \cite{Barsade2002}, thus reinforcing existing health inequalities.\\
\item Our work reveals \textbf{four \emph{emotional tensions}} that arise \textbf{within and between stakeholders} when addressing patients’ emotional needs and other demands relevant for successfully designing and implementing RPM. The emotional tensions we identified are: making diverse patients feel heard vs. exploited (T1), \emph{making them feel seen vs. deprioritized for efficiency} (T2), \emph{making them feel empowered vs. anxious} (T3), and \emph{making patients feel cared for vs. detached from care} (T4). In line with research on empathy and human-centered design, we find that close collaboration between users and designers is important but can be challenging, and that participation is not always without negative consequences for diverse populations (e.g., unintended power dynamics, lack of empathy in understanding users’ needs) \cite{Bennett2019,Lindsay2012,Wright2008}. A common solution proposed to deal with such power dynamics is to arrange meetings where people in higher power (such as doctors) are not present, or talk to specific groups separately (e.g., elderly patients, doctors, designers). \textbf{Our findings thus go beyond existing work by moving away from a siloed to an integrative approach.} Given that RPM is a highly collaborative effort that affects a variety of stakeholders’ interests \cite{Berg1999,Richardson2022}, paying attention to these competing demands is critical to successfully integrate patients’ emotional needs across design phases. Particularly in the context of technology design and implementation processes, neglecting patients’ negative emotions can have detrimental downstream consequences for attaining health equity objectives \cite{Huy2011}.\\
\item We offer \textbf{specific design recommendations for the collaborative design of health technologies} based on \textbf{adopting a \emph{paradox mindset}} where stakeholders search for synergies and see them as opportunities. Such a paradox mindset \cite{Lewis2000,Lewis2014} embraces the inherent emotional tensions, and, instead of looking for ``either-or"' solutions, stakeholders search for ``both-and"' solutions, thereby balancing diverse emotional needs of patients with other demands along the design and implementation process. This includes examples such as moving forward with multiple digital prototypes or balancing digital with in-person care. Previous research has shown that, even in situations of resource scarcity, a paradox mindset helps to balance tensions and fosters innovation, agility, and learning, all critical processes for the inclusive design of equity-focused digital technologies \cite{Mecklai2021}.
\end{itemize}

\subsection{Practical Recommendations for Designing and Implementing RPM}

Our analysis sheds light on how to apply the revealed insights for successfully designing and implementing RPM and similar collaborative technologies and social support systems to foster digital health equity. To manage the previously-neglected \emph{emotional tensions} within and across stakeholders that our study revealed, we suggest adopting a \emph{paradox mindset} to develop ``both-and"' solutions and balance the tensions instead of prioritizing one aspect over another. 
\begin{itemize}
\item \textbf{Tension T1: Making Diverse Patients Feel Heard versus Exploited during Design Phase:} When aiming to access rich stories from patients and making them feel heard, we recommend that designers, operational staff, and providers stay attentive to small cues (changes in voice, posture of patients etc.) that can signal patients’ negative emotions, and that they carefully adapt their questions to the situation and accept the possibility that diverse participants may not wish to share certain aspects of their past and current experience. By behaving with sensitivity, power dynamics and socially desirable response patterns are reduced, which facilitates the access to rich and ``real"' stories without overwhelming patients.
\item \textbf{Tension T2: Making Diverse Patients Feel Seen vs. Deprioritized for Efficiency:} Our findings encourage all stakeholders involved in the design of new technologies to embrace experimentation. Given that the discovery of needs, particularly of diverse patient groups, is challenging even within a single patient group \cite{Duque2019,Lindsay2012,Lindsay2012a} moving forward with multiple ideas is more likely to lead to discovery of equitable and synergistic solutions that might fulfill the needs of the diverse stakeholders involved in the design. To balance the emotional needs of feeling seen with the demand for efficiency, we suggest including all relevant stakeholders (patients, designers, operational staff, and providers) early in the design phase so that they all have a chance to understand the patient’s emotional needs. Alternatively, a possible model for accomplishing this is appointing ``boundary spanners"' \cite{Ancona1990,Carlile2004}, a designated role for individuals who balance patients’ emotional demands of feeling seen with respect to their individual needs with efficiency demands throughout the design and implementation process. We further suggest that all stakeholders engage in cross-occupational perspective-taking and sharing of the goals, needs, and the working and social context of other stakeholders. Only if and when diverse stakeholders are open to engaging in discussion and embracing others’ perspectives will solutions that incorporate both efficiency demands and equity across stakeholder groups be identified \cite{Huber2010,Ren2011}.
\item \textbf{Tension T3: Making Diverse Patients Feel Empowered vs. Anxious:} Stakeholders should not only focus on addressing cognitive needs but also work on downregulating diverse patients’ negative emotions around RPM, allowing them to talk through their fears and anxieties with designers, operational support staff, and providers. To make patients feel empowered to use RPM and overcome their anxieties, healthcare staff members should empathize with patients that they are not alone in their unique challenges and reassure them that RPM use will become more intuitive with time. Deliberate roles and resources to provide for such coaching needs will need to be allocated.
\item \textbf{Tension T4: Making Diverse Patients Feel Cared For vs. Detached from Care:} In order to allow diverse patients to feel deeply cared for, face-to-face contact via video or in-person visits with their provider are critical. Given that previous studies have shown that relational closeness is fostered through continuity of care \cite{Mayer1995}, having one doctor interacting with the same patient over time would likely be beneficial.
\end{itemize}

\subsection{Limitations and Future Research Avenues}

Although our findings significantly contribute to a better understanding of the emotional tensions arising in the design and implementation of RPM technologies, we acknowledge several limitations and boundary constraints. Our qualitative design did not allow us to assess causality. This stream of research will benefit from accessing the emotional needs of patients through quantitative surveys with larger sample sizes and controlled experiments to validate the generalizability and representativeness of our findings and to test the link between specific design strategies and patients’ anxieties and frustration with RPM.

Next, future work on the risks associated with patients’ feeling detached from their care through technology-mediated patient-provider relationships. Future research should explore how intersectionality across diversity dimensions affects diverse patients’ experience when using RPM, and what types of support would benefit groups who are affected by more than one attribute associated with marginalization.

One interviewee openly expressed the increased risks involved for diverse groups, such as elderly patients, who often unquestioningly trust clinicians. At the same time, diverse groups also tend to be more skeptical of healthcare staff members \cite{LaVeist2000}. It therefore seems critical to explore how shifts in agency, such as taking a more active stance in their own healthcare, might lead to diverse patients losing or increasing trust in providers and which ethical considerations must be integrated into RPM to protect diverse patients.

To conclude, identifying emotional needs of diverse patients in the design and implementation of digital health technologies is critical to foster health equity. These emotional needs can be at odds with other stakeholders’ needs and result in tensions. Adopting a paradox mindset where synergistic solutions across stakeholders are explored is key to balance the emotional tensions and successfully design and implement RPM for diverse patients.

\section*{Acknowledgement}

We appreciate funding from the National Science Foundation (Grant number: 1928614 and 2129076) and Swiss National Science Foundation (Grant number: P500PS\_202955 and P5R5PS\_217714).

\bibliographystyle{ACM-Reference-Format}
\bibliography{literature}

\end{document}